\documentclass[a4paper]{jpconf}
\usepackage{amsmath,amsthm,amssymb,amsfonts}
\usepackage{array}
\usepackage{verbatim}
\usepackage[square,sort&compress]{natbib}

\begin{document}

\newcommand{\eg}{\emph{e.g.}\;}
\newcommand{\ie}{\emph{i.e.}}
\newcommand{\nn}{\nonumber}
\newcommand{\Dslash}{\not{\hbox{\kern-4pt $D$}}}
\newcommand{\Supertwistor}{\mathbb{C} \mathrm{P}^{3|4}}
\newcommand{\Twistorspace}{\mathbb{C} \mathrm{P}^{3}}
\newcommand{\MHV}{\mathbb{C} \mathrm{P}^{1}}
\newcommand{\dinstanton}{\mathbb{C} \mathrm{P}^{1}}
\newcommand{\AdS}{\mathrm{AdS}}
\newcommand{\half}{\frac{1}{2}}
\newcommand{\diff}{\mathrm{d}}
\newcommand{\ra}{\rightarrow}
\newcommand{\Zset}{{\mathbb Z}}
\newcommand{\Cset}{{\,\,{{{^{_{\pmb{\mid}}}}\kern-.47em{\mathrm C}}}}}
\newcommand{\gra}{\alpha}
\newcommand{\grb}{\beta}
\newcommand{\grl}{\lambda}
\newcommand{\gre}{\epsilon}
\newcommand{\zb}{{\bar{z}}}
\newcommand{\mn}{{\mu\nu}}
\newcommand{\Acal}{{\mathcal A}}
\newcommand{\Rcal}{{\mathcal R}}
\newcommand{\Dcal}{{\mathcal D}}
\newcommand{\Mcal}{{\mathcal M}}
\newcommand{\Ncal}{{\mathcal N}}
\newcommand{\Kcal}{\mathcal{K}}
\newcommand{\Lcal}{{\mathcal L}}
\newcommand{\Scal}{{\mathcal S}}
\newcommand{\Wcal}{{\mathcal W}}
\newcommand{\Bcal}{\mathcal{B}}
\newcommand{\Ccal}{\mathcal{C}}
\newcommand{\Jcal}{\mathcal{J}}
\newcommand{\Vcal}{\mathcal{V}}
\newcommand{\Ocal}{\mathcal{O}}
\newcommand{\Qcal}{\mathcal{Q}}
\newcommand{\Zcal}{\mathcal{Z}}
\newcommand{\Zb}{\overline{Z}}
\newcommand{\Urm}{{\mathrm U}}
\newcommand{\Srm}{{\mathrm S}}
\newcommand{\SO}{\mathrm{SO}}
\newcommand{\Sp}{\mathrm{Sp}}
\newcommand{\SU}{\mathrm{SU}}
\newcommand{\SL}{\mathrm{SL}}
\newcommand{\U}{\mathrm{U}}
\newcommand{\PSU}{\mathrm{PSU}}
\newcommand{\be}{\begin{equation}}
\newcommand{\ee}{\end{equation}}
\newcommand{\bea}{\begin{eqnarray}}
\newcommand{\eea}{\end{eqnarray}}
\newcommand{\tQ}{\tilde{Q}}
\newcommand{\trho}{\tilde{\rho}}
\newcommand{\tphi}{\tilde{\phi}}
\newcommand{\lt}{\tilde{\lambda}}
\newcommand{\dagphi}{{\phi^\dagger}}
\newcommand{\dagq}{{q^\dagger}}
\newcommand{\dagz}{{z^\dagger}}
\newcommand{\bzeta}{{\bar{\zeta}}}
\newcommand{\blambda}{{\bar{\lambda}}}
\newcommand{\bchi}{{\bar{\chi}}}
\newcommand{\tmu}{\tilde{\mu}}
\newcommand{\mut}{\tilde{\mu}}
\newcommand{\dbar}{\bar{\partial}}
\newcommand{\Comment}[1]{{}}
\newcommand{\doublet}[2]{\left(\begin{array}{c}#1\\#2\end{array}\right)}
\newcommand{\twobytwo}[4]{\left(\begin{array}{cc} #1&#2\\#3&#4\end{array}\right)}
\newcommand{\note}[2]{{\footnotesize [{\sc #1}}---{\footnotesize   #2]}}
\newcommand{\Dc}{\mathrm{D}_c}
\newcommand{\Df}{\mathrm{D}_f}
\newcommand{\Zbar}{\bar{Z}}
\newcommand{\p}{\partial}
\newcommand{\lan}{\langle}
\newcommand{\ran}{\rangle}

\rightline{QMUL-PH-07-21} \rightline{CERN-PH-TH/2007-156}
\rightline{TIFR/TH/07-23}\vspace{-1truecm}

\title{Adding flavour to twistor strings}

\author{J Bedford$^{1,2,a}$, C Papageorgakis$^{3,b}$ and K Zoubos$^{1,c}$}

\address{${}^1$ Centre for Research in String Theory, Department of Physics\\
Queen Mary, University of London, Mile End Road, London E1 4NS, UK}
\address{${}^2$ Department of Physics, CERN - Theory Division\\
1211 Geneva 23, Switzerland}
\address{${}^3$ Department of Theoretical Physics, Tata Institute of Fundamental Research\\ Homi Bhabha Road, Mumbai 400 005, India}

\ead{$^a$james.bedford@cern.ch $^b$costis@theory.tifr.res.in $^c$k.zoubos@qmul.ac.uk}

\begin{abstract}
Twistor string theory is known to describe a wide variety of field theories at tree--level
and has proved extremely useful in making substantial progress in perturbative gauge theory. We explore the
twistor dual description of a class of $\Ncal=2$ UV--finite super--Yang--Mills theories with fundamental flavour by
adding `flavour' branes to the topological B--model on super-twistor space and comment on the appearance of these
objects. Evidence for the correspondence is provided by matching amplitudes on both sides.
\end{abstract}

\section{Introduction}

One of the original motivations for Witten's twistor string theory \cite{Witten0312} was to explain the
simplicity of the Parke--Taylor formula for scattering of gluons at tree--level, which in spinor helicity
variables -- where a light--like momentum $p$ is decomposed in terms of commuting bosonic spinors as
$p_{\gra\dot{\gra}}=\lambda_{\gra}\tilde{\grl}_{\dot{\gra}}$ -- takes the compact form:
\be
\label{parke-taylor}
A(r^-,s^-)=\frac{\lan r\, s\ran^4}{\lan 1\,2\ran\lan 2\,3\ran\ldots\lan n\!-\!1\, n\ran\lan n\,1\ran}\ ,
\ee
where all momenta are taken to be outgoing.
This describes the scattering of any number of positive helicity gluons with two negative helicity gluons
labelled by $r$ and $s$ and the angle brackets denote a scalar product of positive chirality spinors
$\lan i\,j\ran\equiv\lan\lambda_i\,\lambda_j\ran\sim\epsilon_{\gra\grb}\lambda_i^{\gra}\,\lambda_j^{\grb}$.
As such, the holomorphicity of (\ref{parke-taylor}) -- \ie\, its dependence solely on positive chirality spinors --
was a crucial ingredient in the construction of a perturbative string dual to $\Ncal=4$ super--Yang--Mills
(SYM) and thence in understanding the fundamental importance of the so-called maximally helicity violating (MHV)
amplitudes described classically by (\ref{parke-taylor}).

This holomorphicity implies that under a Penrose transform \cite{Penrose67}, whereby one transforms
$\tilde{\grl}\rightarrow i\partial/\partial\mu$ and
$\mu\rightarrow -i\partial/\partial\tilde{\grl}$, (\ref{parke-taylor}) localises on a simple
algebraic curve in the space spanned by $Z^m=(\grl^{\gra},\mu^{\dot{\gra}})$. This algebraic curve is of
degree one and genus zero and is a copy of $\dinstanton$, holomorphically embedded in $\Twistorspace$, the twistor
space of complexified Minkowski space. With the addition of four fermionic directions $\psi^I$, the
target space becomes the Calabi-Yau supermanifold $\Supertwistor$ which in \cite{Witten0312} Witten considered as a
background for the topological open-string B--model.

When supplemented with space--filling branes, the B--model on a Calabi-Yau (CY) \cite{Witten9207} gives rise to
a $(0,1)$--form $\Acal=\diff \bar{Z}^{\bar{m}} \Acal_{\bar{m}}$ whose target space interactions can be encoded by the
cubic holomorphic Chern--Simons theory Lagrangean (written with the help of the CY $(3,0)$--form $\mathbf{\Omega}$)
\be \label{HCSorig}
\mathcal{L}=\textstyle{\frac{1}{2}}\mathbf{\Omega}\wedge\textrm{Tr}
(\Acal\cdot\bar{\partial}\Acal+\textstyle{\frac{2}{3}}\Acal\wedge\Acal\wedge\Acal)\, .
\ee
Similarly, for a Calabi-Yau supermanifold the action is again given by holomorphic Chern--Simons, but
in this case the field $\Acal$ is a superfield depending on the super-coordinates $\psi$ and $\bar{\psi}$. In the
specific case of $\Supertwistor$, Witten considered Neumann boundary conditions on all directions except for the
antiholomorphic $\bar{\psi}$ (`D5'--branes) which leads to a superfield
\be \label{N=4 superfield}
\Acal=A+\psi^I\lambda_I+\textstyle{\frac{1}{2!}}\psi^I\psi^J\phi_{IJ}+
\textstyle{\frac{1}{3!}}\epsilon_{IJKL}\psi^I\psi^J\psi^K\tilde{\lambda}^L+
\textstyle{\frac{1}{4!}}\epsilon_{IJKL}\psi^I\psi^J\psi^K\psi^LG\ ,
\ee
giving the spectrum of $\Ncal=4$ SYM theory.

In this description, the holomorphic Chern--Simons action is defined
on a space with 6 real bosonic dimensions (which also has fermionic coordinates $\psi^I$), and when
transformed to four dimensions, the interactions encoded in (\ref{HCSorig}) actually correspond to those
of self--dual $\Ncal=4$ Yang--Mills rather than those of the full theory. At first sight one thus finds the
spectrum of $\Ncal=4$ SYM but only a subset of the interactions. However, building on an idea due to
Nair \cite{Nair88}, Witten demonstrated that the full set of interactions arise non--perturbatively by
coupling the theory to D1--instantons (Euclidean 2--branes) \cite{Witten0312}. In the case of the MHV
amplitudes these wrap degree one genus zero curves holomorphically embedded in $\Supertwistor$ via the
equations
\be \label{embedding}
\mu_{\dot \alpha} + x_{\alpha\dot \alpha} \lambda^\alpha = 0 \quad
\textrm{and} \quad \psi^I +\theta_\alpha^I \lambda^\alpha = 0\;,
\ee
where the moduli
$x_{\alpha\dot \alpha}=\sigma^{\mu}_{\alpha\dot{\alpha}}x_{\mu}$ and $\theta^I_\alpha$ correspond to the coordinates of
4d Minkowski space and (on--shell) $\Ncal=4$ superspace respectively. The prescription for the
calculation of tree--level MHV amplitudes, and therefore integration over degree
one, genus zero curves,
is then
\be\label{amp}
A_{(n)} = g^2 \int d^4x\; d^8\theta \;\langle \int_{\MHV}\!\!
J_1w_1\cdots \int_{\MHV}\!\! J_nw_n \rangle\;,
\ee
where $J_i$ are D1 worldvolume free--fermion currents coupling to
the external `D5'--brane fields, while the $w_i$'s are the
twistor space equivalents of wavefunctions for the external particles.

This description of weakly-coupled $\Ncal=4$ Yang--Mills inspired a great deal of
progress in the understanding of perturbative gauge theory. In particular, the extension
of (\ref{amp}) to a number of non--MHV tree--level amplitudes was understood in
\cite{Roibanetal0402,RoibanVolovich0402,Roibanetal0403} and a very important related
development was the method of Cachazo, Svr\v{c}ek and Witten (CSW) \cite{Cachazoetal0403} for
using MHV amplitudes as effective vertices in the calculation of tree--level non--MHV processes.
Despite the appearance of conformal supergravity at one loop in twistor string theory
\cite{BerkovitsWitten04}, substantial field theory progress has also been made at the
quantum level as a direct result of \cite{Witten0312}. This is not limited to amplitudes
in maximally supersymmetric Yang--Mills \cite{Brandhuberetal0407} where the CSW rules
have been shown to hold at one--loop, but extends to theories with less or no supersymmetry
\cite{Bedfordetal04,QuigleyRozali04,Bedfordetal0412,Badgeretal07} and even $\Ncal=8$ supergravity \cite{NastiTravaglini07}.
These and other developments are reviewed in \cite{CachazoSvrcek05} where further references can also be found.

\section{$\Ncal=2$ theories with fundamental matter}

The results previously mentioned strongly suggest that twistor string theory is both more widely
applicable than is currently known \emph{and} that its validity should extend into the
quantum regime. As such, and as an intermediate step towards the latter goal, it is interesting
and important to map out the range of four--dimensional theories that can potentially admit a twistor
string description. In this vein, the $\Ncal=1$ exactly marginal deformations of $\Ncal=4$ SYM were
shown to have a twistor dual in \cite{KulaxiziZoubos04}, while various quiver gauge theories arising
as $\Ncal=1$ and $\Ncal=2$ orbifolds of $\Ncal=4$ SYM were treated in \cite{Giombietal04,ParkRey04}.
References to other developments can be found in \emph{e.g.} \cite{BPZ07}.

In continuation of this programme, the present authors showed that it is possible
to use twistor string theory to describe theories with fundamental matter. In particular, the twistor
duals of $\Ncal=2$ SYM with gauge group $\Sp(N)$ and one antisymmetric and four fundamental hypermultiplets and
of $\Ncal=2$ SYM with $\SU(N)$ gauge group and $2N$ fundamental hypermultiplets were identified in \cite{BPZ07} and it is to those
constructions that we now turn. Let us begin with the first of the two aforementioned theories, which we term the
$N_f=4$ theory for brevity. The second ($N_f=2N$) theory is very similar -- in fact simpler in some
senses -- and as such we will only mention it briefly in what follows and refer the reader to \cite{BPZ07}
for further details.

An important part of the identification of a twistor dual (in the sense of \cite{Witten0312}) to a theory
is the ability to split its spacetime Lagrangean into a self--dual part and a part of $\Ocal(g^2)$ which is obtained
as a perturbation about the self--dual theory. This can indeed be done for the $N_f=4$ theory by performing
various helicity dependent rescalings of the fields in the problem. These are detailed in \cite{BPZ07} as
well as a number of field redefinitions, after which (introducing an anti--self--dual two--form $G$ to write the
Yang--Mills action in first--order form) the Lagrangean is:
\bea\label{Nf4full}
\scriptstyle{\mathcal L\ \  =}  & \!\!\scriptstyle{\mathrm{Tr}}&\!\!\!\!\! \scriptstyle{\left[ -\frac{1}{2}  GF +\frac{1}{4}g^2G^2 +
  D\phi^{\dagger} D\phi + i \bar \lambda^a \bar{\sigma}^{\mu}D_{\mu} \lambda_a
 -   \lambda^a\lambda_a \phi^{\dagger} + 2 g^2
 \bar\lambda^a\bar\lambda_a  \phi  \right]  + \mathrm{Tr} \left[\frac{1}{2}
  D z^a_{\phantom a A}D z^A_{\phantom A a}\right.}\nn\\
      &\scriptstyle{+}& \scriptstyle{\left. i \bar\zeta^A  \bar{\sigma}^{\mu}D_{\mu}
\zeta_A  -  z^a_{\phantom a A}[\lambda_a,
 \zeta^A] - 2 g^2 z^A _{\phantom A a}[\bar\zeta_A , \bar \lambda^a] +
 \zeta^A\zeta_A \phi - 2  g^2 \bar\zeta^A \bar\zeta_A
 \phi^\dagger\right]  + \frac{1}{2}  Dq^{ a}_{\phantom{ a}M}  Dq_{\phantom M a}^M}\nn\\
&  \scriptstyle{-} & \scriptstyle{i \bar \eta_M \bar{\sigma}^{\mu}D_{\mu}
\eta^M    +  q^{  a}_{\phantom{ a}M} \lambda_a \eta^M  - \frac{1}{2}\eta_M\phi\eta^M -2 g^2\left(
   \bar\eta_M \bar{\lambda}^a q^M_{\phantom M  a }   +\frac{1}{2} \bar \eta_M \phi^\dagger\bar\eta^M\right)}\nn\\
&\scriptstyle{+}& \scriptstyle{g^2 \left(-\frac{1}{2} q^a_{\phantom a M}
  \{\phi^\dagger , \phi \} q^M_{\phantom M a} +\frac{1}{4}  q^{
 a}_{\phantom a M}
  [z^b_{\phantom b A},z^A_{\phantom A a}] q^M_{\phantom M b}  \right)
 - \frac{g^2}{8}\left( (q^{ a}_{\phantom a M} q^N_{\phantom N a})(q^{
 b}_{\phantom b N} q^M_{\phantom M b})\right.}\\
  &\scriptstyle{+}& \scriptstyle{\left.(q^{ a }_{\phantom a M} q^{ b}_{\phantom b N} )(q^N_{\phantom N a}
 q^M_{\phantom M b}) \right)- g^2\;   \mathrm{Tr}  \left( \frac{1}{2}[\phi^\dagger, \phi]^2
  + \frac{1}{4} [ z^a_{\phantom a A},z^A_{\phantom A b } ] [z^b_{\phantom b
  B},z^B_{\phantom B a}] + [z^a_{\phantom a A } , \phi][
 \phi^\dagger ,z^A_{\phantom A a}] \right)}\nn \; .
\eea

This theory enjoys a large amount of symmetry apart from its supersymmetry and
conformal invariance. $A,G,\phi,\phi^{\dagger},\grl$ and $\bar{\grl}$ are in the adjoint of the gauge group $\Sp(N)$,
while $z,\zeta$ and $\bar{\zeta}$ are in the irreducible second-rank antisymmetric representation and
$q,\eta$ and $\bar{\eta}$ are in the fundamental.
$M=1,\ldots,8$ is an index of a global $\SO(8)$ flavour symmetry under which the fundamental fields are vectors,
while $a$ and $A$ are fundamental indices of two different $\SU(2)$ groups. The $\SU(2)_a$ symmetry is a subgroup
of the $\Ncal=2$ R-symmetry group while $\SU(2)_A$ is a flavour--like symmetry for the antisymmetric fields;
no other field transforms nontrivially under its action.

The aim, then, is to find a twistor string theory describing the tree--level processes of
(\ref{Nf4full}), which we expect to be described by a holomorphic Chern--Simons--like action
enriched with D1--instantons. The approach we take is similar to that of \emph{e.g.}
\cite{Giombietal04}, where the theories there are engineered by acting on the
twistor dual to $\Ncal=4$ SYM. In particular, an orbifold action on the fermionic coordinates of
$\Supertwistor$ breaks (some) supersymmetry (and in some instances the gauge group) but maintains
conformal invariance, since the bosonic directions remain untouched.

However, it is clear from the ideas of \cite{Giombietal04} and also the construction of the
$N_f=4$ theory in terms of 10d string theory as an orientifold of Type IIB
\cite{Sen9605,Banksetal96,Douglasetal96,Aharonyetal96} that a world--sheet
parity operation will also be required. Thus, we act on the fermionic coordinates of $\Supertwistor$ and
the $\Ncal=4$ superfield (\ref{N=4 superfield}) with the following transformations
\be\label{superorientifold}
(a)\quad  \psi^a\ra \psi^a \ \ ,\ \ \psi^A\ra -\psi^A\ \ ; \quad
(b)\quad \Acal^i_{\;\;j}\ra \Omega^{ik}(\Acal^T)_k^{\;\;l}\Omega_{lj}
=(\Acal^T)^i_{\;\;j}
\equiv \Acal_j^{\;\;i}\;,
\ee
where $\Omega_{2N\times2N}$ is the $\Sp(N)$ invariant tensor, and we have
also split the fermionic directions into $I=a,A$ where $a=1,2$ and $A=3,4$. Requiring $\Acal$ to be invariant under this operation,
it is easy to see that one obtains the following decomposition
\bea\label{VZ}
\nn\hat \Acal  &=& (A+\psi^a\lambda_a+\psi^1\psi^2\phi+\psi^3\psi^4\phi^\dagger
+\epsilon_{cd} \psi^3\psi^4\psi^c\tilde{\lambda}^d+
\psi^1\psi^2\psi^3\psi^4 G)\\
&&+\; \psi^A(
\zeta_A+\psi^a z_{Aa}+\epsilon_{AB}\psi^1\psi^2\tilde{\zeta}^B)\; ,
\eea
where the fields in the first line are symmetric under (\ref{superorientifold}) and
thus in the adjoint of $\Sp(N)$ and those in the second are antisymmetric.

We have thus obtained the adjoint and antisymmetric sectors of the $N_f=4$ theory and now
turn our attention to the fundamentals. By analogy with the IIB string description, it should be
clear that incorporating the fundamental fields will require the introduction of a new object in
twistor space. We implement this by adding a new kind of brane to our configuration which we
 call a `flavour' ($\Df$) brane. As this object is expected to wrap the same number of bosonic
directions (but only the $\psi^a$ fermionic directions) as the `D5' branes already present we will
similarly term the `D5' branes `colour' ($\Dc$) branes. Our boundary conditions on open strings
can thus be summarised in the following table:
\begin{table}[h]
\caption{Boundary conditions for open strings in the B--model
  setup.} \label{Bcs}
\begin{center}
\begin{tabular}{cccc} \br
Direction & $\Dc$--$\Dc$ & $\Dc$--$\Df$ & $\Df$--$\Df$ \\ \mr
$Z$,$\Zbar$ & NN & NN & NN  \\
$\psi^a$ & NN & NN & NN  \\
$\psi^A$ & NN & ND & DD \\
$\bar{\psi}^{\bar{a}}$,$\bar{\psi}^{\bar{A}}$ & DD & DD& DD \\ \br
\end{tabular}
\end{center}
\end{table}

The mathematical requirements imposed by fermionic boundary conditions (b.c.'s) are not as clear or
straightforward as those for their bosonic counterparts. In the bosonic case, the concrete
constraints placed on the worldsheet by Dirichlet or Neumann b.c.'s can in many instances be translated
into a tangible geometrical picture in terms of branes and thus precise constraints in spacetime. In the present
case we are dealing with b.c.'s on fermionic directions primarily from a target space perspective and
as such the constraints to be imposed are not as evident. For example, in \cite{Witten0312} the Dirichlet conditions on the
antiholomorphic fermionic directions were interpreted as setting $\bar{\psi}^{\bar{I}}=0$. However, simply
interpreting the Dirichlet conditions on the \emph{holomorphic} fermionic directions encountered here (as shown in Table 1) as
imposing $\psi^A=0$ does not seem to provide the correct degrees of freedom.

The resolution comes with the realisation that one really wants to apply a fermionic analogue of dimensional
reduction. This is part of a more general question of properly defining sub-supermanifolds of supermanifolds,
some aspects of which have been considered in \cite{Saemann04}, and in our case the constraints involved can
be implemented in terms of a suitable set of integral conditions. We will not discuss these constraints any further
here, but now present the fields resulting from them and refer the reader to \cite{BPZ07} for a more thorough discussion.

The introduction of the new $\Df$ branes mean that the $f$--$f$ strings will give rise to fields living on their
worldvolume, which by comparison with the IIB description of the $N_f=4$ theory in
\cite{Sen9605,Banksetal96,Douglasetal96,Aharonyetal96} is expected to be related (possibly by an analogue of the Penrose
transform) to an 8d theory. As such, we are not especially interested in them and furthermore expect them
to decouple from the low energy dynamics of the $N_f=4$ theory. We therefore do not display their explicit form
(which can be found in \cite{BPZ07}), but instead concentrate on the $\Dc$--$\Df$ and $\Df$--$\Dc$ strings. The appropriate
orientifold--invariant state for these gives rise to a $(0,1)$--form field
\be\label{Qcal}
\textstyle{\Qcal^i_{\;X}=\psi^A Q^i_{AX}
=\psi^A\left(\eta^i_{AX}+ \psi^aq^i_{aAX}+ \psi^1 \psi^2\tilde{\eta}^i_{\; AX}\right)}\; ,
\ee
whose conjugate is related under the orientifold action by the condition
$\Qcal^X_{\;\;i}=\Omega^{XY}\Qcal^j_{\;\;Y}\Omega_{ji}$ and where $i$ is an index of the fundamental
of $\Sp(N)$ while $X$ is in fact an index of $\Sp(2)$. The particular form of $Q$ is not new and
was introduced in \cite{Ferber78,Boelsetal06} to describe superfields in the fundamental representation, though its
derivation here in terms of $\Qcal$ provides a stringy description and a natural
explanation for the fermionic nature of $Q$ which had to be assumed in \cite{Ferber78,Boelsetal06}.

That $X$ should be an $\Sp(2)$ index is perhaps rather surprising, given that the $N_f=4$ theory has
an $\SO(8)$ flavour group which is realised in the 10d string theoretic description via the subgroup
$\SU(4)\times \U(1)\subset \SO(8)$. One might therefore have expected either $\SO(8)$ itself or
$\SU(4)$ to show up. In fact what we find is that twistor string theory seems to favour the
subgroup $\Sp(2)\times \SU(2)_A\subset\SO(8)$. Furthermore, the $\SU(2)$ in question is realised
geometrically as the rotations of the fermionic $\psi^A$ into one another. The group on the
$\Df$ branes is thus $\Sp(2)$ (which is perhaps in some sense
unsurprising given that the colour and flavour branes share the same bosonic directions and we
have an $\Sp(N)$ gauge group arising from the $\Dc$'s) and we have a novel situation where an $\SU(2)$ flavour
subgroup has a geometrical realisation.

Having thus obtained the adjoint, antisymmetric \emph{and} fundamental degrees of freedom we can write down
an analogue of the holomorphic Chern--Simons Lagrangean (\ref{HCSorig}) appropriate to the $N_f=4$ theory:
\be \label{Nf=4TotalAction}
\mathcal{L}\ =\ \textstyle{\frac{1}{2}}\mathbf{\Omega}\wedge\big(\mathrm{Tr}(\hat {\mathcal A}\cdot\bar\partial \hat{ \mathcal A}+
  \textstyle{\frac{2}{3}} \hat{ \mathcal
A}\wedge\hat{\mathcal A}\wedge \hat{ \mathcal A})+ \mathcal Q^{ X}\cdot
\bar \partial
\mathcal Q_{ X}+\mathcal Q^{ X}\wedge \hat{ \mathcal
A} \wedge \mathcal Q_{ X}\big)\; ,
\ee
where $\hat{\Acal}$ is as in (\ref{VZ}) and $\Qcal$ as in (\ref{Qcal}). When expanded into component
form, it is this action which we expect to match that of the self--dual truncation of the $N_f=4$ theory (\emph{i.e.} the
$g\rightarrow 0$ limit of (\ref{Nf4full})) via a suitable non--linear generalisation of the Penrose transform in the
spirit of \cite{PopovSamann04}. When coupled to D1--instantons \`{a} la \cite{Witten0312}, we should then be
able to compute tree--level amplitudes in this theory by integrating over the moduli space of genus zero
curves of appropriate degrees and check this proposed duality.

In \cite{BPZ07}, this is precisely what was done and a number of different amplitudes were compared on both
sides of the duality. There, the `pre--analytic' amplitudes, which have $S=-4$ under $S:\psi^I\rightarrow e^{i\beta}\psi^I$,
were shown to vanish -- trivially so from the string theory perspective as they localise on points in twistor space
implying that the kinematic invariants must vanish, and by direct calculation from the 4d point of view. Furthermore a number
of different analytic amplitudes (\emph{i.e.} those that are MHV or have $S=-8$) at 4 and 5--point were found to agree precisely (up to a
constant universal factor) by applying Witten's prescription essentially unmodified on the string theory side and by calculating
the relevant Feynman diagrams on the gauge theory side. Such amplitudes include ones with adjoint particles, antisymmetrics and
fundamentals both as external and internal states and the results match in all cases. This gives us
confidence that the two theories in question \emph{are} dual to one another at tree--level and moreover the
results of the calculations provide further evidence for the geometric realisation of the $\SU(2)$ flavour subgroup discussed previously.

A similar story emerges for the $N_f=2N$ theory, further details of which can be found in \cite{BPZ07}. As expected, the world--sheet
parity operation is unnecessary in this case and the colour--stripped amplitudes again match
with agreement found up to normalisation. We are thus confident that such
$\Ncal=2$ theories that include fundamental matter have a twistor string description, and as a result this
 provides important information about the scope of validity of twistor string theory. More information is
clearly needed in order to uncover the full (quantum) structure of such descriptions, but this, together with
further progress (perhaps, for example, \cite{MasonSkinner0708}) could be of assistance in achieving this goal.
\ack
J.B. would like to thank the organisers of the EPS HEP conference held in
Manchester, UK, July 19--25 for the opportunity to present this work there. He would also like to acknowledge the support of
a Marie Curie early stage training grant and a Queen Mary studentship.
\newpage
\bibliography{proceedings}
\bibliographystyle{iopart-num}

\end{document}